\documentclass[conference]{IEEEtran}
\usepackage{cite}
\usepackage{amsmath,amssymb,amsfonts}
\usepackage{algorithmic}
\usepackage{graphicx}
\usepackage{textcomp}
\usepackage{xcolor}
\usepackage{subcaption}
\usepackage{balance}
\usepackage{booktabs}

\def\BibTeX{{\rm B\kern-.05em{\sc i\kern-.025em b}\kern-.08em
    T\kern-.1667em\lower.7ex\hbox{E}\kern-.125emX}}

\begin{document}
\title{ACCELERATING WRF I/O PERFORMANCE WITH ADIOS2 AND NETWORK-BASED STREAMING \\}

\makeatletter
\newcommand{\linebreakand}{
  \end{@IEEEauthorhalign}
  \hfill\mbox{}\par
  \mbox{}\hfill\begin{@IEEEauthorhalign}
}
\makeatother

 \author{
   \IEEEauthorblockN{1\textsuperscript{rd} Erick Fredj}
   \IEEEauthorblockA{\textit{Computer Science Department}\\
  	\textit{The Jerusalem College of Technology}\\
  	Jerusalem, Israel \\
  	fredj@jct.ac.il\\
  	\textit{Toga Networks, a Huawei Company}\\
  	Tel Aviv, Israel \\
  	erick.fredj@toganetworks.com
  	}
	\and
   \IEEEauthorblockN{2\textsuperscript{nd} Yann Delorme}
   \IEEEauthorblockA{
     \textit{Toga Networks, a Huawei Company}\\
     Tel Aviv, Israel \\
    }
   \and
   \IEEEauthorblockN{3\textsuperscript{nd} Sameeh Jubran}
   \IEEEauthorblockA{
     \textit{Toga Networks, a Huawei Company}\\
     Tel Aviv, Israel \\
     }
  \and
  \IEEEauthorblockN{4\textsuperscript{nd} Mark Wasserman}
  \IEEEauthorblockA{
     \textit{Toga Networks, a Huawei Company}\\
     Tel Aviv, Israel \\
     }
 \and
 \IEEEauthorblockN{5\textsuperscript{nd} Zhaohui Ding}
 \IEEEauthorblockA{
	\textit{Huawei Technologies Co. Ltd.}\\
	Beijing, China \\
	}
 \and
 \IEEEauthorblockN{6\textsuperscript{nd} Michael Laufer}
	 \IEEEauthorblockA{
		\textit{Toga Networks, a Huawei Company}\\
		Tel Aviv, Israel \\
	}
 }


\maketitle

\begin{abstract}
With the approach of Exascale computing power for large-scale High Performance Computing (HPC) clusters, the gap between compute capabilities and storage systems is growing larger.
This is particularly problematic for the Weather Research and Forecasting Model (WRF), a widely-used HPC application for high-resolution forecasting 
and research that produces sizable datasets, especially when analyzing transient weather phenomena. Despite this issue, the I/O modules within 
WRF have not been updated in the past ten years, resulting in subpar parallel I/O performance.

This research paper demonstrates the positive impact of integrating ADIOS2, a next-generation parallel I/O framework, as a new I/O backend option in WRF.
It goes into detail about the challenges encountered during the integration process and how they were addressed. The resulting I/O times show an over 
tenfold improvement when using ADIOS2 compared to traditional MPI-I/O based solutions. Furthermore, the study highlights the new features available to WRF 
users worldwide, such as the Sustainable Staging Transport (SST) enabling Unified Communication X (UCX) DataTransport, the node-local burst buffer write capabilities 
and in-line lossless compression capabilities of ADIOS2.
	
Additionally, the research shows how ADIOS2's in-situ analysis capabilities can be smoothly integrated with a simple WRF forecasting pipeline, resulting in a 
significant improvement in overall time to solution. This study serves as a reminder to legacy HPC applications that incorporating modern libraries and tools 
can lead to considerable performance enhancements with minimal changes to the core application.
\end{abstract}


\section{Introduction}
The use of High Performance Computing (HPC) in scientific applications is facing an increasing I/O challenge, leading to performance bottlenecks in simulation pipelines, 
such as weather forecasting models. To address this issue, this paper investigates the integration and application of ADIOS2, a high-performance I/O and data management library,
with the widely used HPC application, WRF (Weather Research and Forecasting Model).

\section{Related Work}
\label{section:related}
In the past, various studies have investigated I/O scaling and bottlenecks in WRF. For example, Kyle \cite{akira} and NCAR found that I/O time exceeded compute time at scale 
when running the Hurricane Maria 1km test case on the Cheyenne and Yellowstone supercomputers with more than 2000 compute cores. Balle et al. \cite{balle2016improving} demonstrated 
that write times increased as more nodes were added, with I/O time reaching 50\% of the total run time when node counts reached about 500. 
However, the WRF Quilt Server was successfully used to reduce I/O time at the cost of computational resources. Finkenrath et al. \cite{finkenrath} found that using the PnetCDF option 
resulted in a ten-fold speedup compared to the serial-based NetCDF, and running WRF in hybrid MPI+OpenMP mode greatly reduced I/O time. 
Another study applied the first version of ADIOS to the GRAPES mesoscale Numerical Weather Prediction application \cite{ZOU2014378}, achieving a ten-fold increase in I/O times compared 
to their MPI-I/O approach, but they did not examine in-situ pipelining capabilities. Singhal and Sussman \cite{Singhal} integrated a version of ADIOS into WRF, but it was not 
upstreamed to the WRF community and did not utilize ADIOS2's code coupling and transport capabilities.

Several I/O middleware libraries, such as MPI-I/O  \cite{Corbett95overviewof}, NetCDF \cite{netcdf}, Parallel-NetCDF \cite{Li2003}, and HDF5 \cite{hdf5}, have been introduced, aiming to optimize I/O performance using parallel I/O for 
remote parallel file systems. However, these libraries have limitations and cannot directly take advantage of emerging high-speed node-local storage. 
The present study investigates the integration of the ADIOS2 library, which offers advanced features and can take advantage of emerging high-speed node-local storage. 
The paper is structured as follows: related works are discussed in Section \ref{section:related}; 
the background of WRF and ADIOS2 is presented in Section \ref{section:Background}; implementation details and challenges of integrating a new I/O library into WRF 
are detailed in Section \ref{section:design}; results of a set of I/O performance evaluations comparing ADIOS2 to other available WRF I/O methods, as well as an 
example in-situ analysis pipeline, are discussed in Section \ref{section:Results}; and conclusions and next steps are presented in Section \ref{section:conclusions}.

\section{Background}
\label{section:Background}
The following section details the background of both the WRF model and the ADIOS2 library.

\subsection{Weather Research Forecasting Model}
\subsubsection{WRF Background}
WRF is a cutting-edge mesoscale Numerical Weather Prediction system that is designed for both atmospheric research and forecasting. It is an open-source project 
that has gained popularity worldwide and is officially supported by the National Center for Atmospheric Research (NCAR). The WRF software framework enables 
efficient and massively parallel computation across a wide range of computing platforms, making it a true community model. The model is based on the compressible, 
non-hydrostatic atmospheric motion equations that include multiple physics processes such as cloud and precipitation, boundary layer turbulence, land-ocean air interaction, 
radiative transfer, and energy transfer at the surface. Finite difference method is used to discretize these equations, and the resulting time-dependent atmospheric 
motion and physical states are computed through integration. Due to the large number of prognostic variables in the three dimensions, which is a result of the 
multiple physical processes involved in atmospheric motion, computational and storage resources must be high performance. WRF operates in two phases, the first of which involves 
configuring the model domain, preparing initial conditions, and ingesting input data, while the second runs the forecast model and outputs solution and checkpoint files. 
WRF is predominantly written in Fortran and can be built with a variety of compilers. It runs on platforms with UNIX-like operating systems, from laptops to supercomputers, 
and its software framework handles I/O and parallel-computing communications.

\subsubsection{WRF I/O Backends}
WRF’s well-defined I/O API provides several different implementations of its I/O layer, the ones relevant for the present work:

\begin{itemize}
\item  Serial NetCDF\cite{netcdf} (\emph{io\_form=2}): The default I/O option in the WRF model. When this I/O option is selected, all data is funneled through the first MPI rank, where this rank alone writes out a NetCDF4 based file using the NetCDF library (HDF5 based). While Rank 0 is writing to disk, all other ranks wait until the write has fully concluded before continuing computation. This method performs well at low process counts but at higher counts, the write can quickly dominate the computation time. One of the main advantages of this method is the ability to use lossless compression that is integrated within HDF5. This results in much smaller file sizes, achieving compression ratios close to 4. Still, due to the massive communication overhead, and single write thread, this option achieves poor I/O performance.
\item Split NetCDF (\emph{io\_form=102}): This option also uses the NetCDF library for I/O but instead of sending all data to the first MPI rank, each rank writes its own distinct file. As will be seen later in this work, this method is able to achieve very high throughput at moderate MPI rank counts due to the absence of communication costs, but this file-per-process method (\emph{N-N}) does not scale to high counts due to the immense pressure to the underlying file system and metadata servers. Additionally, as this output method outputs multiple distinct files, the post processing is not trivial, especially when the rank of readers does not match the amount of files. To counter this, a community provided routine can stitch the output files together back into a single file, but this also incurs a non negligible time and resource cost, as well as additional complexity in post processing pipelines.
\item Parallel NetCDF\cite{Li2003} (\emph{io\_form=11}): WRF's primary parallel I/O option that utilizes MPI-I/O. When this method is employed, all MPI ranks cooperate to write a single output file in parallel using PnetCDF, which directly accesses MPI-I/O. As opposed to NetCDF4 based methods, this method does not allow for data compression. Even without compression capabilities, this option has been shown to offer an order of magnitude increase in write bandwidth compared to the Serial NetCDF method at scale, due to  coordinated MPI-IO two-phase method \cite{Weifeng2018}. As this is the primary parallel I/O option that allows for operation without requiring additional overhead from stitching multiple files together, or file format conversions, it is treated as the benchmark method, when comparing against the new ADIOS2 approach.
\item Quilt Servers: The quilt server technique uses dedicated I/O processes ("servers") that deal exclusively with I/O, enabling the compute processes to continue with their work without waiting for data to be written to disk before proceeding. Data from multiple compute ranks are merged ("quilted") together by a dedicated I/O rank by means of MPI communication calls and kept in system memory until they are written to PFS. This was previously found to be high performing I/O option available in WRF, even though compute resources are sacrificed and memory usage can be exceedingly high. This option is not investigated in this work, but should be investigated in future works.
\end{itemize}

\subsection{ADIOS2}
ADIOS2, developed as the successor to the Adaptable Input Output System (ADIOS) by Lofstead et al. \cite{adios}, is a highly flexible library that allows users to configure 
different I/O techniques, file formats, and transports through a simple XML file. This adaptability makes it a suitable option for use across different scales, 
ranging from laptops to supercomputers. Although ADIOS2 is coded in C++, it provides support for various programming languages such as C, Fortran, Python, and Matlab.

ADIOS2 is primarily focused on high-performance file-based I/O, despite offering in-situ analysis and Wide-Area Network (WAN) and UCX transport capabilities. It uses its proprietary file format, 
BP5 \cite{adios2}, to assign specific MPI ranks as aggregators that write sub-files to a BP5 output directory while collecting streaming data from the sub-group ranks. 
The aggregator writes the received data to disk continuously, without encountering file locking issues that MPI-I/O-based approaches like  Parallel NetCDF (PnetCDF) \cite{Li2003} 
and HDF5 \cite{hdf5} often face. The metadata algorithm tracks the location of data buffers within the sub-files to reassemble the data for reading.

ADIOS2 provides tunable aggregators and placement at runtime, and the current version defaults to a single aggregator per node for optimal shared memory communication 
while minimizing the number of processes accessing the underlying filesystem. The combination of sub-files, streaming data, and the absence of global data sharing results 
in significant enhancements in write bandwidth. Moreover, the BP5 file format offers node-local burst buffer support, where each process writes sub-files to its high-speed 
local file system, and the burst buffer data is drained back to the Parallel File System via a separate thread.

ADIOS2 is continuously evolving and has already been incorporated into numerous essential HPC applications, with more support in the pipeline.

\begin{itemize}
	\item OpenFOAM\cite{openfoam}
	\item LAMMPS\cite{lammps}
	\item XGC\cite{XGC}
	\item E3SM:\cite{E3SM}
	\item Trilinos\cite{trilinos}
	\item PETSc\cite{petsc}
\end{itemize}

At runtime, ADIOS2 provides multiple engines, including SST, which allows direct connection between data producers and consumers, bypassing the filesystem and 
using the new UCX communication. SST supports a variable number of readers and writers, buffering data in the producer's memory until the consumer is ready to receive it. 
This feature allows for seamless in-situ postprocessing.

ADIOS2 supports in-line data manipulation, including lossless compression through various compressors and codecs such as the Blosc meta-compressor \cite{blosc}. 
The library is under continuous development and has already been integrated into several essential HPC applications, with more support expected to come.

\subsection{UCX}
UCX, which stands for Unified Communication X, is a high-performance communication library for distributed computing systems that enables efficient data transfer between 
computing nodes in a parallel application. It is designed to provide a unified API for a variety of interconnect technologies, such as InfiniBand, RoCE, Ethernet, and others, 
and supports both point-to-point and collective communication.

UCX is designed to provide low-latency, high-bandwidth communication, and supports a range of communication operations, including send and receive, scatter and gather, 
reduce, allreduce, broadcast, and others. It also supports asynchronous communication and provides a range of features to optimize communication performance, 
such as zero-copy data transfers, message batching, and hardware offload.

UCX achieves high performance by using a number of techniques, including multi-threading, cache-aware algorithms, and optimized memory management. 
It also takes advantage of hardware features such as remote direct memory access (RDMA), which allows data to be transferred directly between the memories of two nodes without 
involving the CPU.

Overall, UCX provides a high-performance, portable, and scalable communication framework that can be used to develop parallel applications for a variety of distributed 
computing systems.

\section{Design and Implementation}
\label{section:design}
The ADIOS2 I/O backend implementation in WRF is similar to other external I/O options, with some tiny differences. 
It substitutes the \emph{ncmpi\_put\_var\_type} calls of PnetCDF with \emph{adios2\_put} calls of ADIOS2, which are easier to use. However, ADIOS2 and NetCDF differ in how they handle the time 
dimension. NetCDF treats time as a separate dimension, 
while ADIOS2 is step-based and places data at each new time step. Thus, the main WRF I/O logic loop was modified to provide ADIOS2 with the start and end step information. 
To avoid editing the XML file for each output variable, the compressor option was added to the namelist.input file, allowing compression to be applied to all variables. 
To increase user adoption, a converter program was developed to convert the ADIOS2 output file back into a NetCDF file. This incurs a time penalty of about 10 seconds 
per time step, but allows the use of existing NetCDF pipelines. However, this tool does not allow for the use of ADIOS2's in-situ analysis pipeline capabilities. 
NCAR has merged the ADIOS2 I/O module and converter script for use by WRF users worldwide as of February 2023.

\section{Results}
\label{section:Results}

In this section, we present an evaluation of the ADIOS2 backend in the WRF model, which includes a comparison with 
other I/O options available in the model. The study focuses on investigating the SST with UCX Data transport, the ADIOS2 
burst buffer write capabilities, and the impact of in-line compression on write times and output data sizes. Additionally, 
we showcase a basic ADIOS2-based in-situ post-processing pipeline and compare its end-to-end run times with traditional post-processing methods.

To conduct the tests, we utilized a compute cluster consisting of 8 nodes, each equipped with two 18-core Intel Xeon Gold 6240 CPUs, 384 GB DDR4 
memory, and a Mellanox ConnectX-6 interconnect. The storage setup involved a dedicated storage node with a BeeGFS file system striped over eight 
10K RPM spinning hard disk drives connected to the compute nodes via Mellanox ConnectX-5 NICs. Additionally, each compute node had an Intel DC 
P4510 1TB NVMe SSD drive.

For testing, we used WRF v4.4 in distributed memory mode (i.e., \emph{dmpar}), along with NetCDF v4.8.1, PnetCDF v1.12.1, and ADIOS2 (Github, master branch) 
compiled using GCC v10.2. The CONUS 2.5km case, a well-known benchmark for WRF, was selected for I/O testing, and we increased the WRF history file frequency 
to one file every 30 minutes output history file to reflect a relevant data analysis time scale during \cite{benchdatav44}.

We performed the tests on up to 8 compute nodes (288 MPI ranks) five times for each configuration and computed the average I/O times. It is worth noting that the 
official WRF benchmarks were incompatible with WRF v4.4, and hence we replaced them with the \emph{New} CONUS 2.5km benchmark developed by Kyle \cite{akira}. 
Our evaluation presents a comprehensive analysis of the ADIOS2 backend within WRF, highlighting its strengths and limitations compared to other I/O 
options available in the model.

\subsection{ADIOS2 File Write}
To evaluate the performance of three different I/O configurations using the BeeGFS file system as the storage target, the new CONUS 2.5km model was employed. 
The first configuration was the baseline parallel I/O option, which used PnetCDF. The second configuration utilized the Split NetCDF method, and the third 
configuration incorporated the new ADIOS2 method. Fig. \ref{fig:adios2_comparison} demonstrates the average write times for the WRF history file for each 
I/O option at different node counts, highlighting the scalability of each method.

The aim of the evaluation was to compare the performance of the three I/O options and determine the most efficient one for the new CONUS 2.5km model. 
The results of the evaluation could help optimize the I/O performance of the model, which is crucial for timely data analysis and decision making.

\begin{figure*}[htbp]
	\centerline{\includegraphics[]{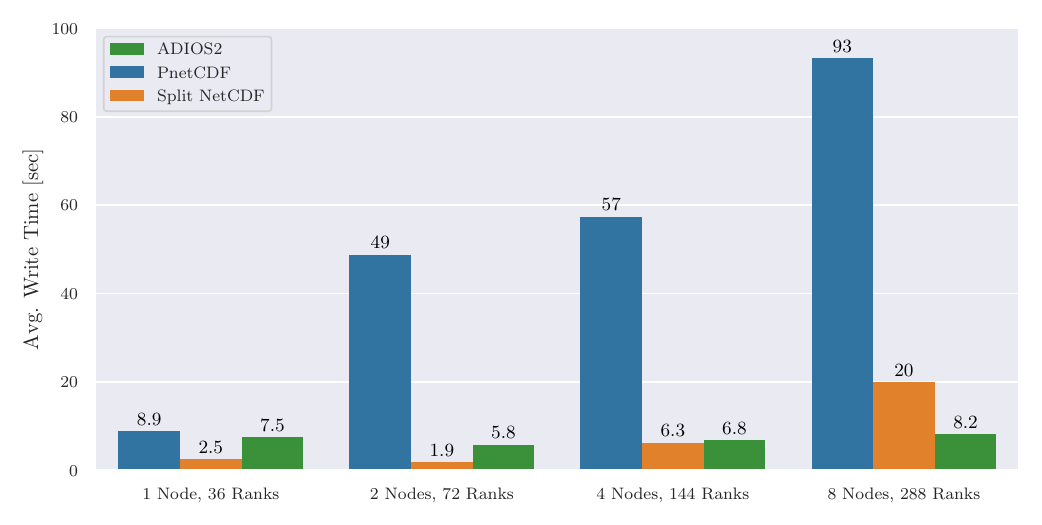}}
	\caption{The graph illustrates a comparison between the average write times of the WRF history file using ADIOS2 and legacy parallel I/O options for different node 
	and rank counts for the CONUS 2.5km model. With 8 nodes and 288 ranks, ADIOS2 demonstrates a more than tenfold improvement over PnetCDF.}
	\label{fig:adios2_comparison}
\end{figure*}

The write times of PnetCDF increase as more nodes are used due to the additional inter-node communication required by the two-phase MPI-I/O based method used in its 
implementation. On the other hand, while the Split NetCDF method shows impressive results at low node counts, its write time increases by a factor of 3 between 4 
and 8 nodes, demonstrating scaling issues inherent in the file-per-process approach.

In contrast, the ADIOS2 method yields the most consistent results across the range of process counts tested, outperforming the PnetCDF results by an order of magnitude 
and halving the write time of the Split NetCDF method when 8 compute nodes are used.

\subsection{ADIOS2 Burst Buffer}
To evaluate the performance of ADIOS2's burst buffer feature, adjustments were made to the \emph{adios2.xml} file, targeting the node-local NVMe SSDs on each compute node. 
This allowed the ADIOS2 aggregators to write data locally, while a background thread continued to drain the burst buffer contents back to the parallel file system (PFS). 
However, the drain feature was disabled for this set of tests.

Fig. \ref{fig:adios2_bb_comparison} illustrates the scaling performance of ADIOS2's burst buffer compared to normal PFS write. At low node counts, the burst buffer results 
exhibit similar times to the PFS write configuration, but as more nodes are added, there is a dramatic decrease in average write time. This is due to the additional 
potential write bandwidth of the node-local NVMe SSDs on each additional node.

To further demonstrate the performance benefits of the ADIOS2 burst buffer feature, the speedup compared to a single node was plotted in Fig. \ref{fig:adios2_bb_scaling}. 
The results showed ideal write time scaling up to 4 nodes, with only a small deviation from ideal at 8 nodes. This is in stark contrast to the inverse speedup trend observed in 
the MPI-I/O based results of PnetCDF as more nodes are added.

Overall, the ADIOS2 burst buffer functionality greatly accelerates I/O write performance. In this case, it showed a significant \emph{two} order of magnitude speedup 
compared to the benchmark WRF parallel I/O option, PnetCDF, on the system used in this work.

\begin{figure}[htbp]
	\centerline{\includegraphics[]{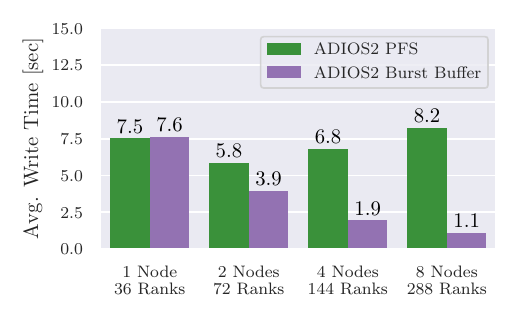}}
	\caption{The figure illustrates the average write times for the WRF history file using ADIOS2 with and without the burst buffer feature on the CONUS 2.5km model. 
	The comparison shows the performance of writing to the node-local burst buffer versus writing to the PFS.}
	\label{fig:adios2_bb_comparison}
\end{figure}

\begin{figure}[htbp]
	\centerline{\includegraphics[]{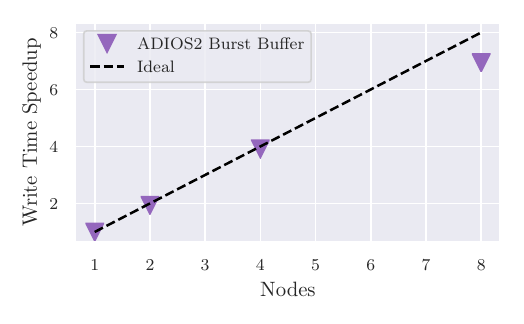}}
	\caption{The plot shows the average speedup of history write time when using the ADIOS2 burst buffer feature compared to a single node, as a function of the 
	number of compute nodes. The results demonstrate that the scaling closely approaches ideal values, with an almost perfect speedup up to 4 nodes and a 
	slight deviation from ideal at 8 nodes.}
	\label{fig:adios2_bb_scaling}
\end{figure}

\subsection{ADIOS2 Aggregator Count}
The ADIOS2 file based I/O is mainly controlled by the number of sub-files written to the file system, which is called the aggregator ratio. By default, 
ADIOS2 writes a single sub-file per node, with one MPI rank serving as the aggregator for the remaining ranks on the node. However, depending on the 
capabilities of the underlying file system, a different number of sub-files/aggregators may be more efficient. ADIOS2 offers the flexibility to adjust 
the number of aggregators at runtime using the \emph{adios2.xml} file or programmatically within the application.

To find the optimal aggregator ratio for the CONUS 2.5km model, a set of tests were conducted where the aggregator ratio was varied. The results, 
shown in Fig. \ref{fig:adios2_aggregator}, indicate that for this system, the ideal aggregator ratio is 36, which corresponds to one aggregator per node 
(each node has 36 cores). When the aggregator ratio was set to 1, where each rank writes its own sub-file, the results were not favorable, similar to the 
file-per-process approach observed in the previous Split NetCDF tests. This finding is consistent with the scaling results seen in the XGC application 
in \cite{adios2}, where the sub-file-per-process approach was not efficient. For all subsequent tests, the number of aggregators/sub-files was set to one per node.

\begin{figure}[htbp]
	\centerline{\includegraphics[]{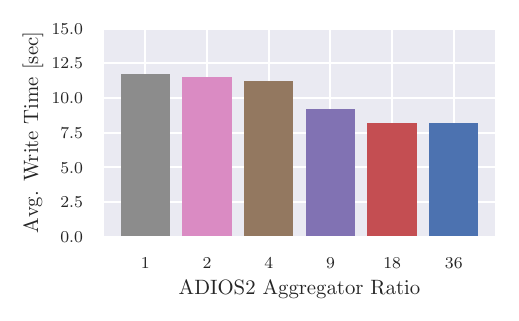}}
	\caption{The impact of varying the ADIOS2 aggregator ratio parameter on the average history write time of the CONUS 2.5km model using 8 compute nodes is shown. 
	The optimal performance is achieved with a single aggregator per node, corresponding to the default ADIOS2 behavior (aggregator ratio of 36 for this cluster).}
	\label{fig:adios2_aggregator}
\end{figure}

\subsection{ADIOS2 in-line Compression}

ADIOS2 offers the Operator abstraction, which allows for in-line data manipulation. One of the primary uses of this abstraction is the ability to compress data in-line. 
This can be accomplished using various available lossy and lossless compression backends and codecs. 
In this study, the Blosc\cite{blosc} "meta-compressor" was chosen as it is lossless and supports multiple 
state-of-the-art compression codes. We tested the following Blosc compression codecs:

\begin{itemize}
	\item BloscLZ\cite{blosc}
	\item LZ4\cite{lz4}
	\item Zlib\cite{zlib}
	\item Zstandard\cite{zstd}
\end{itemize}
Fig. \ref{fig:adios2_compression} findings are particularly significant because they demonstrate that using 
compression with ADIOS2 does not result in any significant performance degradation, even when dealing with 
large-scale parallel I/O. This is important because, in many scientific domains, data sets can be massive, 
and writing them to disk can be a significant bottleneck in the computational workflow. By using compression, 
it is possible to reduce the amount of data that needs to be written, which in turn can lead to significant 
performance improvements.

Furthermore, it is worth noting that the choice of compression codec can have a significant impact on performance. 
In this study, the Zstandard codec was found to be the most effective, which suggests that it may be a good 
choice for other scientific applications that require high-performance parallel I/O. However, it is also possible 
that other codecs may be more effective for specific types of data, and further research is needed to fully 
explore this issue.

Overall, the results presented in Fig. \ref{fig:adios2_compression} suggest that using ADIOS2 compression can
be an effective way to improve the performance of parallel I/O, particularly when dealing with large data sets. 
By choosing the right compression codec, it may be possible to achieve even greater performance improvements, 
which could have significant implications for a wide range of scientific applications.

\begin{figure*}[htbp]
	\centerline{\includegraphics[]{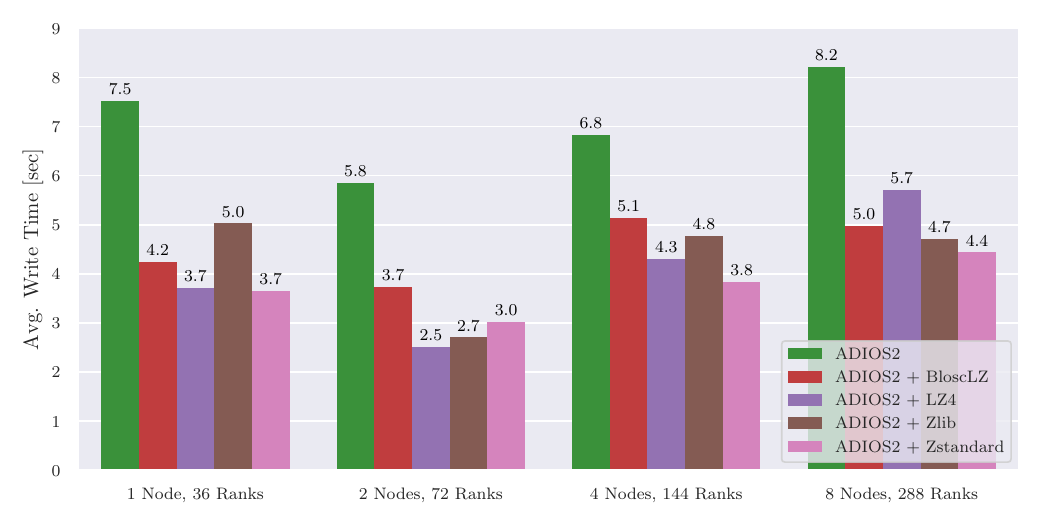}}
	\caption{This figure compares the average history file write times of ADIOS2 compressed data using various 
	Blosc compression codecs to the write times of uncompressed ADIOS2 data. The results indicate that using 
	ADIOS2 compression leads to a nearly 50\% reduction in average write time compared to the uncompressed 
	configuration.}
	\label{fig:adios2_compression}
\end{figure*}

The results from Fig. \ref{fig:adios2_compression_size} demonstrate that both ADIOS2 (Blosc) compression 
and the NetCDF4-based compression methods achieve compression ratios of about 4, leading to significant 
reductions in I/O server contention and required storage volume. Furthermore, the Zstandard codec outperforms 
other Blosc codecs, with the smallest file size and maximal throughput, except for Zlib. These findings 
indicate that Zstandard is an excellent choice for WRF as a premier compressor codec option. Therefore, 
it has been selected as the default codec once compression is enabled in the WRF-ADIOS2 I/O backend.

\begin{figure}[htbp]
	\centerline{\includegraphics[]{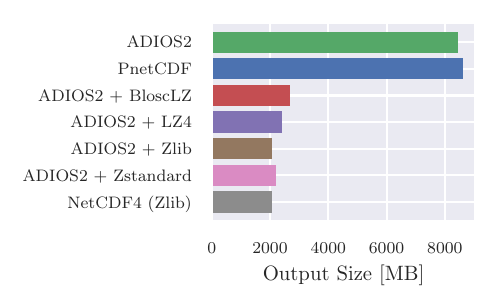}}
	\caption{Comparison of the compressed vs uncompressed output data size using different ADIOS2 compression codecs. }
	\label{fig:adios2_compression_size}
\end{figure}

\subsection{Ideal ADIOS2 File Write Configuration for WRF}
The test run conducted on 8 nodes using the optimized ADIOS2 configuration demonstrated significant improvements in I/O time. 
The use of node-local NVMe SSDs as the ADIOS2 storage target, along with the Blosc compressor and Zstandard codec, 
led to a perceived I/O time of around half a second. This is a significant improvement compared to the I/O 
bottleneck observed when using the PnetCDF I/O method. The results of the optimization are summarized 
in Table \ref{tab:progression}.
\begin{table}[h!]
	\caption{Progression of Optimizations.}
	\centering
	\begin{tabular}{|l|c|c|}
	\hline
	Configuration & Write Time [s] & Speedup\\
	\hline
	PnetCDF             & 93                 & 1X      \\
	ADIOS2              & 8.2                & 11X     \\
	ADIOS2+BB           & 1.1                & 84X     \\
	ADIOS2+BB+Zstandard & 0.52               & 179X    \\ \bottomrule
	\hline
	\end{tabular}
	\label{tab:progression}
\end{table}

As can be seen, the I/O bottleneck observed at the beginning while using the PnetCDF I/O method 
is virtually eliminated when using the ideal ADIOS2 configuration, as the perceived I/O time 
within the application falls to approximately half a second.

\subsection{ADIOS2 In-situ Post-Processing}

ADIOS2 is not just an I/O library, but also a data management library, and its in-situ analysis 
capabilities were tested in a simple weather forecasting pipeline. The pipeline was set up for 
a 2-hour forecasting run, with a history file output every 30 simulation minutes. The aim of 
the test was to show the significant decrease in total time-to-solution when using an in-situ 
pipeline with ADIOS2, as compared to the standard process-after-run method with the benchmark PnetCDF.

For this test, the ADIOS2 SST engine was selected using the \emph{adios2.xml} file, 
which buffers and transfers requested data to a consumer over the network, rather than writing it 
out to a file on the file system. To perform post-processing, a Python-based analysis script was 
written, which plotted a slice of the temperature field over the continental United States and 
generated an image similar to the one in Fig. \ref{fig:pipeline}. Two versions of the script 
were created, one that used the netcdf4-python to read data, and one that used the ADIOS2 
high-level Python API. It's worth noting that the ADIOS2-based script didn't need to be modified 
to support in-situ processing, as the support is inherent when using the stepping mode with the 
Pythonic \emph{for fstep in adios2\_fh} directive.

Fig. \ref{fig:runtime} shows the runtime progression of the ADIOS2-based end-to-end pipeline, 
compared to the PnetCDF sequential pipeline. The I/O, compute, and initialization times were 
extracted and analyzed from the WRF \emph{rsl.out} output files. The ADIOS2-based pipeline 
results showed an almost constant block of compute, as the perceived write time by the application 
was below one second for each of the outputs. The SST engine internally buffered the data and sent it 
to be processed in parallel while the computation continued. On the other hand, with the PnetCDF 
pipeline, the computation was stopped for long periods during the writing process. After the 
computation was completed, the PnetCDF post-processing script was run, which further increased 
the time-to-solution.

In total, the in-situ ADIOS2 approach using the SST engine was able to almost halve the time-to-solution 
compared to the legacy PnetCDF-based approach, providing significant value for time-sensitive applications.

\begin{figure*}[htbp]
	\centerline{\includegraphics[width=500pt,keepaspectratio]{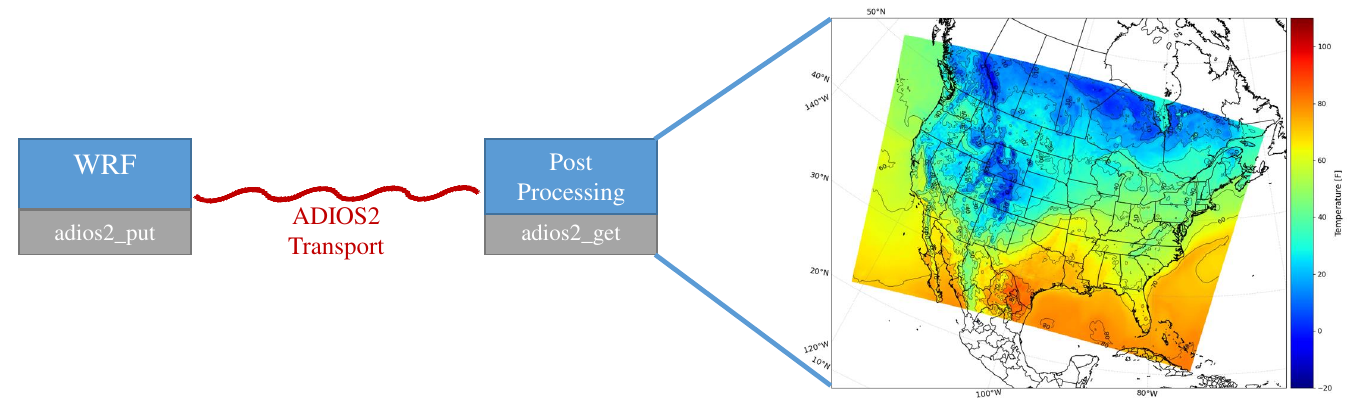}}
	\caption{Run time comparison of a WRF run with postprocessing. The ADIOS2 configuration processes the output data in-situ, using data streamed from WRF, while the PnetCDF configuration uses the traditional process-after-job-completion approach. }
	\label{fig:pipeline}
\end{figure*}

\begin{figure*}[htbp]
	\centerline{\includegraphics[width=500pt,keepaspectratio]{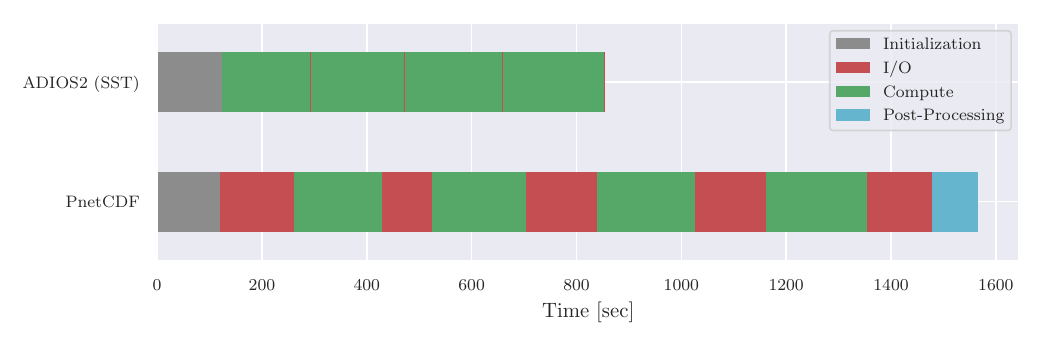}}
	\caption{Run time comparison of a WRF run with postprocessing. The ADIOS2 configuration processes the output data in-situ, using data streamed from WRF, while the PnetCDF configuration uses the traditional process-after-job-completion approach. }
	\label{fig:runtime}
\end{figure*}

\subsection{UCX network-based streaming over BP file-based}
Network-based streaming and file-based streaming are two different methods of delivering content over the internet.
Network-based streaming delivers content in real-time as shown in Fig. \ref{fig:adios2_ucx_bp4_bp5}, 
where the content is continuously delivered as a stream of data, rather than as a complete file that needs to be downloaded 
before rerun can begin. File-based streaming, on the other hand, requires the entire file to be downloaded before rerun can begin.

Here are some benefits of network-based streaming over file-based streaming:

\begin{itemize}
\item Faster rerun: Network-based streaming allows for faster run as the content is delivered in real-time. Users can begin 
streaming the content almost instantly, without having to wait for the entire file to download.

\item Less storage space: Network-based streaming doesn't require users to download and store large files on their devices, 
which can be especially beneficial for users with limited storage space.

\item Smooth run: Network-based streaming can adjust the quality of the stream based on the user's internet connection, 
ensuring that the playback is smooth and uninterrupted.

\item Live streaming: Network-based streaming can deliver live content, such as live weather forecast events, in real-time. 
This allows users to watch events as they happen, rather than having to wait for the file to be made available for download.

\item Security: Network-based streaming can provide better security, as content is not stored locally on the user's device, 
making it less vulnerable to theft or loss. Additionally, network-based streaming can implement secure communication 
protocols such as HTTPS or encryption to ensure that the content is delivered securely.
\end{itemize}

\begin{figure}[htbp]
	\centerline{\includegraphics[width=250pt,keepaspectratio]{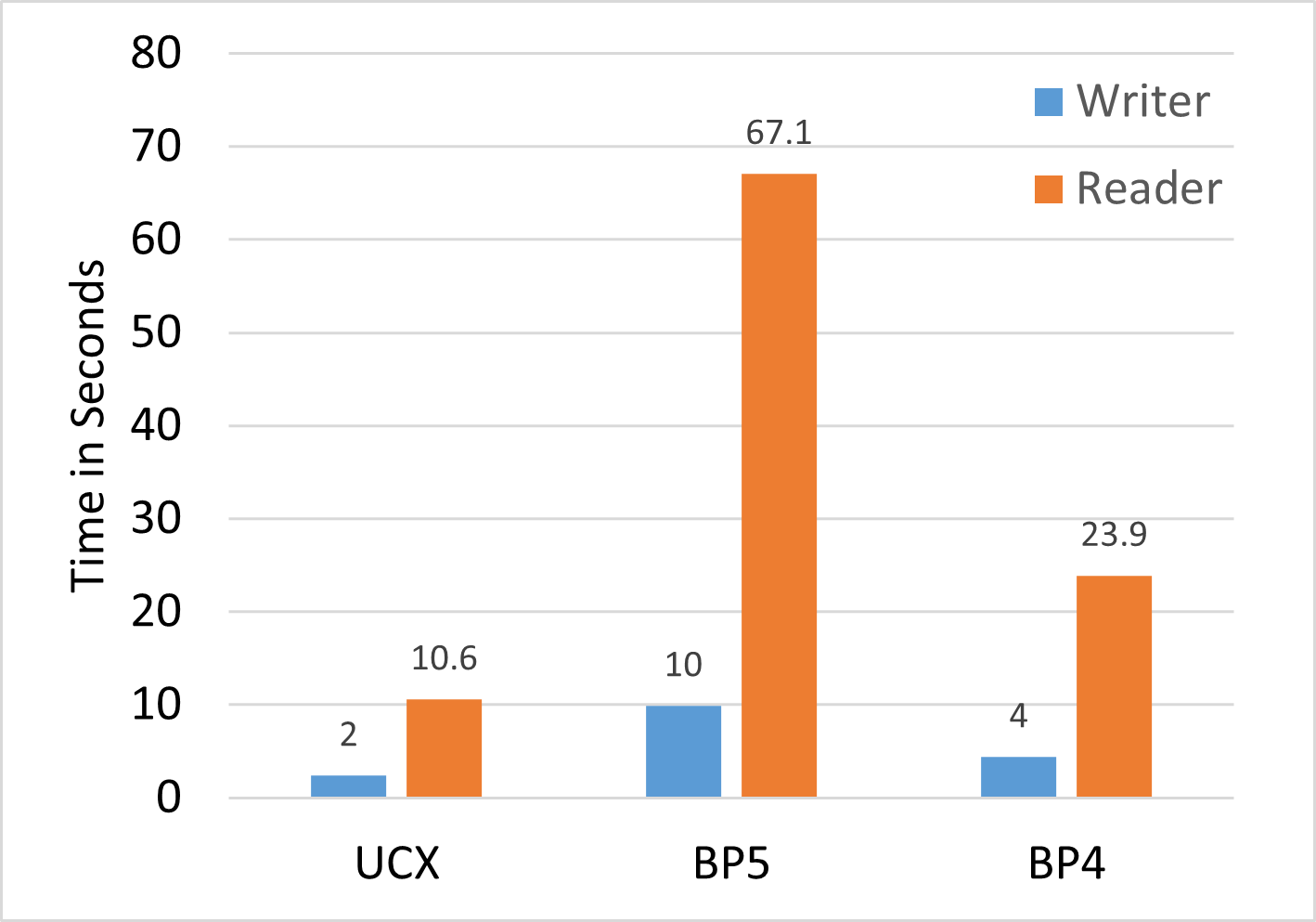}}
	\caption{ A WRF run with postprocessing is compared in terms of run time. The graph shows that the ADIOS streaming engine 
	outperforms the file-based ADIOS2 engine BP4 and BP5 by a factor of 2X-6X on both the Reader and Writer sides.}
	\label{fig:adios2_ucx_bp4_bp5}
\end{figure}

Overall, network-based streaming provides a faster, more seamless, and secure experience for users, making it a 
preferred method of content delivery over file-based streaming.

\subsection{ADIOS2 In-Situ Enabling UCX data transport}

The development of sustainable and efficient transport engines is crucial for the advancement of high-performance 
computing and data management. One such engine, the Sustainable Staging Transport (SST), has recently undergone 
a significant upgrade to enhance its capabilities.

In the first step of this upgrade, the SST engine was fortified by incorporating the UCX communication framework 
as a Remote Direct Memory Access (RDMA)-capable dataplane. This incorporation has expanded the capabilities of 
the ADIOS2 (Adaptable Input/Output System) and has improved hardware compatibility. The UCX transport, which is 
a low-level communication library, has provided a more efficient and flexible interface to manage data 
communication between different nodes.

The incorporation of the UCX communication framework as a RDMA-capable dataplane in the Sustainable Staging Transport
engine has led to significant performance improvements see Fig. \ref{fig:adios2_ucx_bp4_bp5}. Specifically, the UCX transport has outperformed the 
traditional transport middleware layer of EVPATH, resulting in faster data transfer rates and a more 
seamless user experience.

This improvement in performance can be seen in Fig. \ref{fig:adios2_sst_ucx}, which demonstrates the superiority 
of the UCX transport over EVPATH in terms of both throughput and latency. The graph clearly shows that the UCX 
transport is capable of handling a much higher volume of data and is also able to transmit data with much lower 
latency than EVPATH.

\begin{figure}[htbp]
	\centerline{\includegraphics[width=250pt,keepaspectratio]{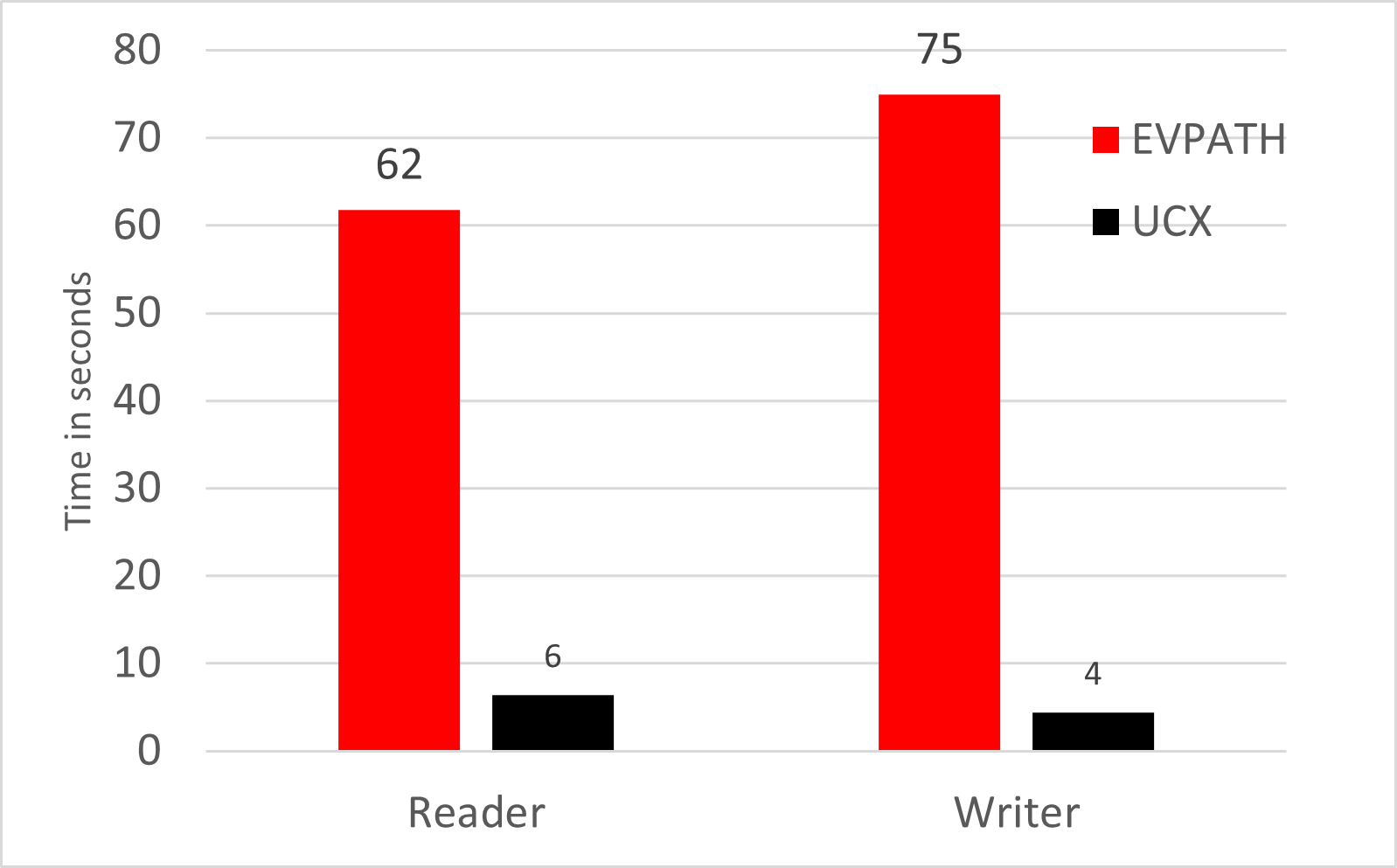}}
	\caption{ A WRF run with postprocessing is compared in terms of run time with 4 nodes. The ADIOS2 SST engine with UCX data transport transport (black) outperformed the default EVPATH data transport (red) by an order of magnitude, with QueueLimit=1.}
	\label{fig:adios2_sst_ucx}
\end{figure}

Overall, the integration of the UCX transport into the SST engine has greatly enhanced its capabilities and has made 
it more efficient and adaptable for data management and high-performance computing. With this upgrade, users can now 
experience faster data transfer rates and a more seamless experience, ensuring that the SST engine remains a valuable 
tool for researchers and scientists.
As a result of this upgrade, the new UCX SST dataplane has been successfully integrated into ADIOS2 2.9, 
enabling users to take advantage of the advanced features and improved performance. The incorporation of the UCX 
transport has made the ADIOS2 engine more adaptable, efficient, and future-proof, ensuring that it remains a 
valuable asset for data management and high-performance computing.

In ADIOS2, the QueueLimit parameter is used with the SST transport method to control the maximum number 
of items that can be queued for asynchronous data transfers between application and engine.
When using the SST transport method, the QueueLimit parameter is set using the $adios2::transport::SSTParams$ class.
The parameter QueueLimit sets the maximum number of items that can be queued for asynchronous data transfers.
This value can be changed according to the available resources and the size of the data being transferred.
According to Fig. \ref{fig:adios2_sst_queuelimit}, setting the queue limit parameter to 0 {\it "asynchronous"} improved UCX optimization, 
while setting it to 1 {\it "synchronous"} improved EVPATH data transfer optimization. As a result, the queue limit parameter 
significantly influenced the optimization of both UCX and EVPATH Data Transfer.

\begin{figure}[htbp]
	\centerline{\includegraphics[width=250pt,keepaspectratio]{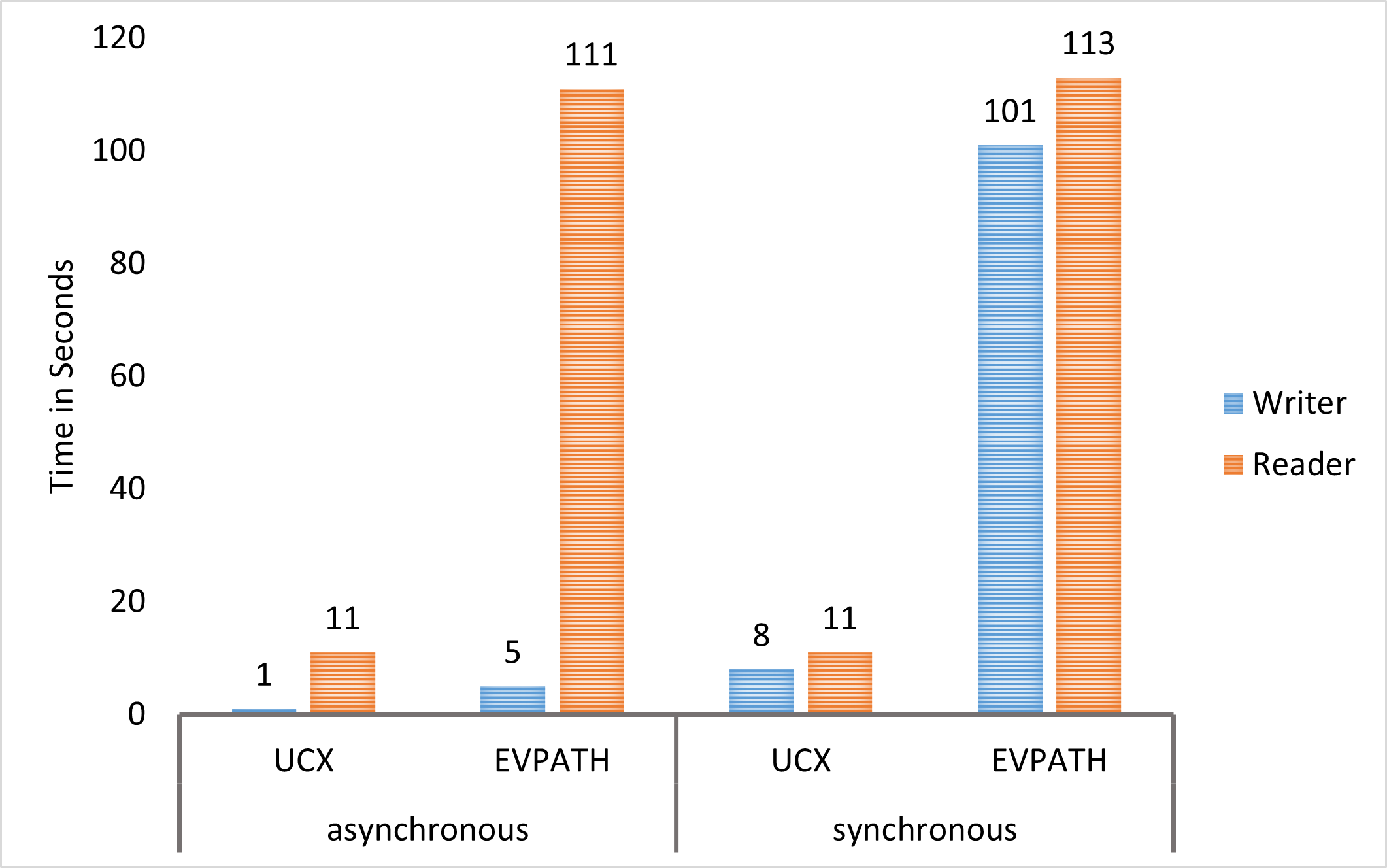}}
	\caption{ Run time comparison of a WRF run with postprocessing with 8 nodes. 
	The ADIOS2 SST engine asynchronous data transfers queuelimit parameter (QueueLimit=0 {\it asynchronous}, QueueLimit=1 {\it synchronous}) improved both middleware 
	layer UCX  and EVPATH  by an order of magnitude.}
	\label{fig:adios2_sst_queuelimit} 
\end{figure}

\section{Conclusions}
\label{section:conclusions}

The ADIOS2 data management library has been integrated into WRF, and this work has demonstrated the implementation 
and notable performance improvements. In addition to accelerating typical PFS writes, the new library also offers 
a wide range of capabilities, like node-local burst buffer write support, high-performance lossless and lossy 
compression, potential two-way data coupling, and in-situ analysis. When using the new I/O backend at scale, 
test findings in this work demonstrate a one to two order of magnitude improvement in perceived write time 
within WRF. Also, it was demonstrated that, when compared to the old PnetCDF technique at scale, 
a sample weather forecasting pipeline using the ADIOS2 SST engine cut the overall time to solution in half. 
Using the ADIOS2 SST engine with UCX data transport, an additional half-time overall solution was achieved.

The benchmark CONUS 2.5km example was used to test the new features and performance improvements of the ADIOS2 
library within WRF. Future work will make use of these new data streaming capabilities to address large 
and research-relevant WRF simulation scenarios that are currently stymied and immobilized by slow, 
sequential analytical pipelines and poor I/O performance. It is also necessary to investigate the 
impact of using lossy compression methods for numerical weather prediction. It is critical to carefully 
balance the increase in effective I/O throughput that can be attributed to the lossy compression codecs 
included in ADIOS2 against the loss in numerical precision.

The results of this work demonstrate the enormous benefits gained by combining cutting-edge open source libraries 
such as ADIOS2 with legacy HPC applications such as WRF, demonstrating how bottlenecks that emerge over time 
(such as the I/O bottleneck in WRF) can be squashed by using the right set of tools.

\section{Acknowledgments}
The authors thank Dr. Swati Sighal from the Department of Computer Science University of Maryland for her assistance 
in this study.

\balance
\bibliographystyle{IEEEtran}
\bibliography{IEEEabrv,./copernicus_2023}
\end{document}